\documentclass[conference]{IEEEtran}
\IEEEoverridecommandlockouts

\usepackage{cite}
\usepackage{amsmath,amssymb,amsfonts}
\usepackage{algorithmic}
\usepackage{graphicx}
\usepackage{textcomp}
\usepackage{xcolor}
\def\BibTeX{{\rm B\kern-.05em{\sc i\kern-.025em b}\kern-.08em
    T\kern-.1667em\lower.7ex\hbox{E}\kern-.125emX}}
\begin{document}

\title{
Towards User Preference Alignment in LLM Recommendation via Explicit Context Feedback 
}

\author{
\IEEEauthorblockN{
Weizhi Zhang\textsuperscript{1,2 *}, 
Wooseong Yang\textsuperscript{1 *}, 
Yuxin Cui\textsuperscript{2}, 
Zhaohui Guo\textsuperscript{2}, 
Hins Hu\textsuperscript{2}, 
Liangwei Yang\textsuperscript{1}, 
Henry Peng Zou\textsuperscript{1}, \\
Qifei Wang\textsuperscript{2}, 
Hanqing Zeng\textsuperscript{2}, 
Jiayi Liu\textsuperscript{2}, 
Yinglong Xia\textsuperscript{2}, 
Philip S. Yu\textsuperscript{1}
}
\IEEEauthorblockA{
\textsuperscript{1}University of Illinois Chicago, Chicago, USA \\
\textsuperscript{2}Meta, Menlo Park, USA \\
Emails: wzhan42@uic.edu
}
\thanks{\textsuperscript{*}Contributed equally.}
}

\maketitle

\begin{abstract}
Traditional recommender systems (RecSys) primarily infer user preferences from implicit signals (such as clicks, watches, and purchases), often neglecting the rich explicit contextual feedback users provide through verbal text, like comments and reviews. This explicit context feedback captures the nuanced reasons behind user decisions regarding their preferences. In addition, it offers critical heterogeneous information for user preference alignment and more explainable recommendations. Overlooking such signals can lead to misaligned user preferences and further reinforce filter bubbles, as algorithms fail to understand the “semantic context” behind user choices. Recent advances in Large Language Models (LLMs) present new opportunities to harness user-generated content for more accurate and diverse recommendations, yet current LLM-based recommendations still focus on using item meta-data and underutilize this resource. In this paper, we advocate for prioritizing explicit context feedback in the next generation of LLM-based RecSys. We review the evolution of recommendation paradigms, highlight the value of context-rich feedback, call for new benchmarks and metrics, and introduce frameworks for integrating explicit user signals into scalable LLM-driven RecSys. Centering on user-preference modeling, we aim to foster more personalized, transparent, and explainable RecSys online platforms.
\end{abstract}


\section{Introduction}
Recommender systems (RecSys) have traditionally learned user preferences from implicit signals, such as clicks, views, or purchases, while often overlook the rich contextual cues embedded in users’ explicit verbal feedback, including reviews, comments, and textual critiques. This explicit context feedback encapsulates the nuanced reasons behind a user’s likes and dislikes, information vital for fine-grained user preference alignment. For example, a user might purchase a product many times but comment “slightly sweet for my taste”, explicitly signaling a dislike for excessive sweetness. Such fine-grained opinions are usually lost in conventional RecSys since they only capture the implicit signals as the reward and optimization goal. 
Recent advances in review-based recommendation have highlighted the value of modeling textual feedback to uncover the latent aspects influencing user opinions~\cite{chen2015recommender, hasan2024based}. By analyzing natural-language reviews, these methods reveal preference structures that numerical ratings or browsing logs fail to capture, leading to more interpretable and user-aligned recommendations. However, despite their promise, most review-based approaches \cite{chen2015recommender, shuai2022review, xiong2021counterfactual} remain non-LLM systems, relying on traditional text encoders or aspect extraction techniques that lack the deep semantic understanding now achievable with large language models (LLMs).

Ignoring these explicit signals can lead to misaligned recommendations and contribute to filter bubbles. When algorithms only learn from past implicit behavior, they risk over-personalizing content and reinforcing a narrow set of interests. Users have reported that their feeds become repetitive echoes of past interactions~\cite{li2025beyond}. In response, users try to provide explicit negative feedback (e.g., sending user feedback to corresponding platforms or writing critical comments regarding the items). However, these behaviors are often ignored by existing industrial RecSys. This oversight highlights a critical gap: recommender systems need to actively leverage explicit, context-rich feedback to achieve alignment with users’ true preferences and promote diversity and novelty beyond historical behavioral patterns. In other words, contextual feedback is key to preference alignment, as it enables systems to understand not only what a user engages with but also the reasons underlying their behavioral patterns.

Large Language Models (LLMs), with their capacity to understand natural language, offer a timely opportunity to bridge this gap by comprehending user behaviors and aligning user preference \cite{deng2025onerec} for recommendation~\cite{hua2023tutorial}. However, even many LLM-based recommenders so far have mostly focused on item metadata or learned from interaction logs, without deeply integrating the wealth of user-authored context~\cite{bang2025llm}. Notably, \cite{wang2025user} introduces a scalable framework that aligns recommendation generation with implicit user feedback signals such as clicks, dwell time, and playback completion. However, despite being a pioneering industrial deployment of LLM-based user feedback alignment, this framework relies exclusively on implicit behavioral data, without incorporating any explicit verbal or textual feedback from users. This leaves significant room for improvement by incorporating such explicit context feedback.

\begin{figure*}[!h] 
    \centering
    \includegraphics[width=0.7\linewidth]{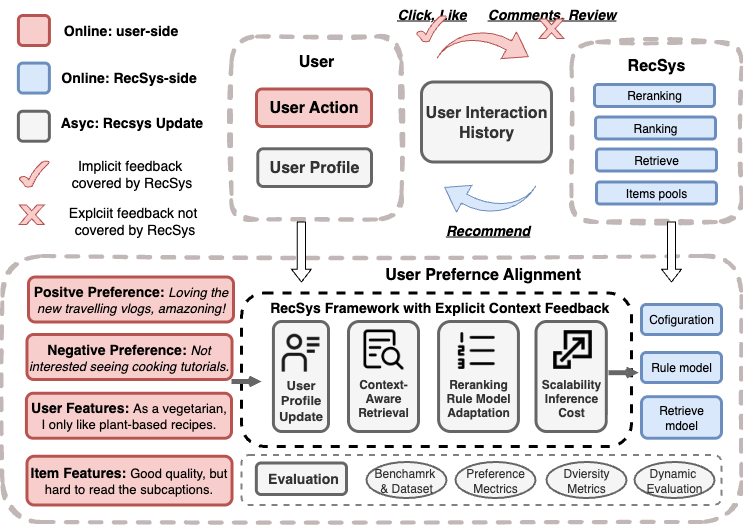}
    \caption{Limitations of existing recommender systems (primarily rely on implicit feedback while overlooking explicit contextual feedback) and framework of LLM-based RecSys with explicit contextual feedback, which incorporates explicit contextual feedback to achieve fine-grained user preference alignment through user profile construction, context-aware retrieval, adaptive reranking, and scalable trade-offs.
    }
\label{fig: framework}
\end{figure*}

In this visionary paper, we advocate for explicit context feedback as a priority research domain in the next generation of recommender systems in the LLM era. We first review the landscape of traditional recommendation paradigms (Section II), then illustrate through examples how user-provided context expresses nuanced tastes (Section III). We propose new directions for benchmarks and metrics to evaluate context-aware preference alignment (Section IV). As shown in Figure~\ref{fig: framework}, we categorize four types of explicit feedback (Section V) and present a forward-looking framework for integrating these signals into LLM-based recommendation (Section VI). Our central thesis is that by aligning recommendations with the textual voice of the user, their own words about their preferences, we can achieve more personalized, transparent, and diverse recommendation outcomes, ultimately improving user satisfaction and trust in the systems.

\vspace{10pt}
\section{Existing RecSys Paradigms with Implicit Interaction Feedback}

\subsection{Content-based RecSys}

Content-based recommendation builds user and item profiles based on descriptive features, then matches users to items with desired attributes \cite{pazzani2007content,javed2021review}. This paradigm leverages item metadata and user attributes, such as textual descriptions, categorical tags, or contextual features, to recommend content semantically similar to what a user has previously engaged with.
Modern content-based architectures extend this principle through neural feature encoders, forming a family of deep content models. 
Representative examples include Wide\&Deep \cite{cheng2016wide}, DeepFM \cite{guo2017deepfm}, DCN \cite{wang2017deep}, and Two-Tower (Dual-Encoder) architectures \cite{huang2013learning}. These models jointly learn nonlinear feature interactions and dense embeddings for both users and items, enabling them to generalize beyond manual or sparse feature engineering.
Despite architectural differences, all maintain a content-driven paradigm, relying on user and item attributes rather than collective user behaviors.
These models excel in cold-start and data-sparse scenarios and offer interpretable, feature-based personalization \cite{zhang2025cold}. 

Although neural models reduce manual feature design, they still rely on domain-specific feature pipelines (text embeddings, categorical encodings, normalization). This confines users to a narrow set of familiar items, often referred to as the “filter bubble” effect, thereby hindering serendipitous discovery of novel or diverse content \cite{kaminskas2016diversity, ross2022echo}. In addition, the “content” used in these models (e.g., demographic or static textual features) often misses behavioral signals such as sequential actions, user patterns, or temporal intent shifts.

\vspace{5pt}
\subsection{Collaborative Filtering and Graph-Based RecSys}
Collaborative Filtering (CF) \cite{koren2021advances, zhang2024mixed, aljunid2025collaborative} operates on the principle of homophily, the idea that users who agreed in the past will agree in the future, by leveraging community-wide interaction patterns. Early memory-based methods focused on finding neighborhoods of similar users or items to make predictions \cite{yu2004probabilistic}. The paradigm saw a significant leap with model-based techniques like Matrix Factorization (MF), which learns latent feature vectors for users and items to capture their underlying characteristics \cite{aggarwal2016model,bokde2015matrix}. This was further advanced by Neural Collaborative Filtering (NCF) \cite{he2017neural}, which employs deep neural networks to model more complex, non-linear user-item relationships. More recently, graph-based methods have gained prominence by representing interactions as a user-item bipartite graph. Techniques using graph neural networks \cite{wu2020comprehensive, ma2023graph}, such as LightGCN \cite{he2020lightgcn}, learn user and item embeddings by aggregating information from their multi-hop neighbors, capturing the higher-order connectivity that MF-based models often miss. Follow-up works develop more advanced self-supervised learning paradigms \cite{wu2021self, yu2022graph, zhang2025sgcl} or graph convolution methods \cite{mao2021ultragcn, zhang2024we, xu2023graph, xu2025graph} to model the user-item interaction patterns.

However, these approaches are fundamentally limited by their reliance on interaction data, which captures the presence of an interaction but not the underlying rationale behind it. A primary issue is the ambiguity of ratings and interactions due to user bias, where the same score or behavior from different users can signify different preference levels \cite{chen2023bias}. These models also struggle with the cold-start problem for new users and items, and can amplify popularity bias, trapping users in filter bubbles \cite{zhang2025cold,klimashevskaia2024survey}. These limitations stem from a common failure: the inability to process the nuanced signals found in explicit context feedback during the graph convolution process. Incorporating such feedback could clarify user rationale, thus enabling more genuine preference modeling.

\subsection{Sequential and LLM-based RecSys}

The evolution beyond static models led to sequential recommendation, which captures the dynamics of user preferences by modeling the chronological order of interactions \cite{wang2019sequential}. While early methods used Markov Chains, the field was revolutionized by deep learning, with Recurrent Neural Networks (RNNs) and later Transformer-based architectures becoming the state-of-the-art for learning a user's intent from their recent behavior \cite{rendle2010factorizing,hidasi2015session,kang2018self,sun2019bert4rec}. The latest paradigm shift involves LLMs, which are leveraged for their immense world knowledge and language understanding. Current applications typically use LLMs as powerful content encoders for item metadata, as advanced sequence models for next-item prediction, or as zero-shot recommenders that generate suggestions from a prompted interaction history \cite{zhao2024recommender,li2023text,geng2022recommendation,hou2024large}.

While the progression from small sequential models to large foundation systems has greatly improved the ability to capture complex behavioral patterns, a fundamental limitation persists. Both paradigms are still largely confined to interpreting implicit signals; sequential models can predict a user's next action without understanding their motivation, and most LLM-based applications apply their power to the same interaction logs rather than to user-generated text. This underutilization of the models' core comprehension capabilities prevents recommendations from being grounded in user rationale. This highlights a critical gap that explicit context feedback is uniquely positioned to fill, as processing a user's own words allows systems to move beyond behavioral patterns to comprehend stated preferences, thus enabling a higher fidelity of preference alignment.

\section{Nuanced Preferences via Explicit Context Feedback: Examples}

Users often express their preferences in their own words, providing clues far more revealing than any click, watch, purchase, or star rating alone. Consider a few intuitive scenarios:

\begin{itemize}

\item \textbf{Example 1: Personal Taste beyond a General Category.}
A user repeatedly dislikes horror films but adds comments like ``Too many jump scares, I prefer psychological horror."
Implicit-only models may overgeneralize, avoiding the entire horror genre. However, explicit feedback clarifies the reason: the user dislikes jump scares, not horror itself. With this insight, the system can recommend suspenseful, thought-driven thrillers instead of gore-heavy titles. This prevents over-filtering and diversifies exposure while maintaining relevance. Explicit context therefore mitigates filter bubbles and enables calibrated diversification.

\item \textbf{Example 2: Maximize User Satisfaction.}
A user rates a smartphone 4 stars and comments, ``The screen is stunning, but the battery drains too fast.'' While a conventional recommendation model only treat this as positive engagement, a context-aware system recognizes ways to improve: the user values display quality but dislikes poor battery life. This allows the RecSys to refine personalization by emphasizing devices with long battery life and excellent screens, avoiding the same flaw that prompted dissatisfaction. Explicit aspect feedback thus interprets coarse ratings into fine-grained preferences to maximize user satisfaction, even given positive feedback signals.

\item \textbf{Example 3: Long-Term Values and Identity Cues}.
A restaurant reviewer writes, “As a vegetarian, I love their plant-based options.”
This short statement expresses a stable personal attribute of vegetarianism that should influence recommendations across domains (e.g., groceries, recipes, dining). Similarly, a news reader might comment, “I prefer human-written pieces over AI-generated ones,” revealing a value-based constraint. Traditional RecSys miss such signals because they rely solely on implicit behaviors. Capturing these identity cues allows RecSys to respect user values and build trust beyond user modeling.

\end{itemize}

These examples collectively highlight that explicit context feedback captures the why behind user actions, revealing preferences, tastes, and values that implicit signals alone cannot express. By analyzing users’ natural language feedback, RecSys can identify aspect-level sentiments (e.g., “great screen but poor battery”), nuanced genre preferences (e.g., “psychological horror over gore”), and enduring personal traits (e.g., “vegetarian” or “prefers human-written content”). Such signals enable fine-grained preference modeling, promote calibrated diversity, and ensure alignment with user identity and intent. Integrating explicit feedback thus enhances both accuracy and transparency: recommendations become explainable (“You mentioned you dislike jump scares, so we avoided them”) and more trustworthy. Ultimately, explicit context feedback transforms one-dimensional behavioral data into rich, interpretable user profiles, driving personalized, value-aligned, and user-controllable recommendation experiences.

\section{Benchmarking Context-Aware Feedback and New Evaluation Metrics}
Note that many existing explicit context feedback signals are stored but not integrated into RecSys, and we need to rethink how we evaluate recommendation performance and design new benchmarks that stress-test a model’s ability to handle such feedback. Traditional benchmarks (MovieLens~\cite{harper2015movielens}, Amazon Review~\cite{hou2024bridging}, etc.) emphasize accuracy of predicting held-out interactions or ratings. Success is measured with metrics like RMSE, Recall, and NDCG, which don’t directly capture alignment with user explicit preferences. We propose several directions for benchmarking and metrics to push the field toward context-aware feedback modeling.

\subsection{Feedback-Enriched Datasets}
New benchmark datasets should include user textual feedback alongside interactions. For example, modern media platforms generate streaming textual context, such as real-time “barrage” comments during movie playback. An augmented MovieLens where each rating is coupled with the user’s verbal feedback or final review.  In this case, a model’s task could be to predict not only what the user will interact with, but also to demonstrate understanding of their verbal feedback (e.g., update the user profiles over the context happened in the movie, based on real-time user comments). One concrete direction is a continuous, sequential recommendation task with user feedback, where the system is fed a timeline: at each step a new user feedback arrives, and the model must update its recommendations accordingly. This reflects real-world usage where user preferences shift as they provide more feedback.

\subsection{Evaluation of Preference Alignment} Beyond traditional ranking-based evaluation pipeline, we need metrics that measure preference alignment: how well recommendations agree with the user’s stated likes/dislikes. One direction is a “dislike avoidance rate”, measuring the fraction of recommended items that do not contain any attributes the user explicitly dislikes. Conversely, a “preference fulfillment score” could measure how many of the user’s explicitly mentioned preferred aspects appear in the recommended items. For example, if a user says “I love movies with surprise twist endings,” does the system’s top-N recommendations predominantly include movies known for twist endings? These metrics require metadata or manual labeling of items with those aspects, or the use of an LLM to classify items’ features, which is now feasible. They directly reward systems for satisfying user-specified criteria.

\subsection{Diversity and Dynamic Preference Adaptation} Explicit context feedback fosters diversity, helping RecSys break information cocoons and filter bubbles. When recommendations rely solely on past implicit signals, users are trapped in narrow content loops that reinforce existing preferences. In contrast, real user behavior and preference are dynamic, and explicit textual feedback reveals such evolving tastes and contextual cues. This dynamic adjustment introduces novel yet relevant items that sustain curiosity and prevent recommendation fatigue. Evaluating diversity~\cite{adomavicius2011improving}, novelty~\cite{hurley2011novelty}, and temporal adaptability thus becomes essential for ensuring that recommendation systems remain both engaging and aligned with users’ changing interests over time.

\section{Taxonomy of Explicit Context Feedback}
Not all explicit feedback conveys the same kind of information. Users express preferences through diverse linguistic signals that differ in intent, granularity, and scope. We categorize explicit context feedback into four major types: positive preferences, negative constraints, user-specific attributes, and item-level evaluations, each offering distinct insights into user intentions and enabling complementary pathways for personalized and explainable recommendations.

\subsection{Preferred and Positive Signals}

Preferred or positive signals are explicit cues where users directly express favor to certain aspects or features. A hotel review stating “Loved the cleanliness and friendly staff” conveys precise, high-fidelity information that implicit signals such as clicks cannot provide. Review-aware models such as DeepCoNN~\cite{zheng2017joint} and NARRE~\cite{chen2018neural} have demonstrated that learning from textual praise improves recommendation accuracy and interpretability. More recent LLM-based approaches, including P5~\cite{geng2022recommendation} and profile management frameworks~\cite{bang2025llm}, illustrate how positive feedback can be distilled into structured user profiles.

Future research should elevate positive feedback from auxiliary evidence to a primary supervisory signal. Dynamic aspect distillation, for example, can leverage LLMs to extract praised attributes and incrementally update user–aspect embeddings. Similarly, contrastive alignment methods~\cite{v2025contrastive,yang2024item,wang2024rdrec} can anchor positive textual cues to pull aspect-rich items closer in embedding space, enhancing both personalization and diversity. Beyond alignment, positive feedback can also promote exploration: when users praise an unfamiliar genre, the system can generalize this as diversification signals, guiding future recommendations toward novel yet relevant items.

\subsection{Dislikes and Negative Signals}

Negative feedback defines the boundaries of user preference that they might choose to leave the platforms, providing explicit constraints on what to avoid. Statements such as “Terrible battery life” or “Too sweet for my taste” serve as strong supervisory cues that implicit non-engagement cannot capture. Aspect-based sentiment analysis~\cite{liu2022sentiment} and neural review models~\cite{zheng2017joint,chen2018neural} show how textual complaints refine user representations. More recent contrastive approaches~\cite{wei2021contrastive} explicitly use disliked items as counterexamples to sharpen embedding separation.

Looking forward, negative signals should support adaptive debiasing and preference calibration. Constructing negative preference graphs, where user–aspect links represent aversions (e.g., “no jump scares”), can complement positive graphs to form complete rationale profiles. LLMs can further distinguish soft versus hard constraints, weighting them accordingly in retrieval or ranking. Dynamic constraint learning~\cite{wang2023learning} allows systems to adjust recommendations in real time to respect user restrictions while maintaining diversity. In this sense, negative signals function not merely as corrective data but as protective alignment layers, ensuring recommendations remain both precise and personalized.

\subsection{User-Specific Attributes and Contextual Factors}

User-specific feedback often reveals information about the user's long-term attributes, such as identity, background, or short-term situational context. These self-declared attributes enable fine-grained personalization beyond demographic inference. Context-aware frameworks have long incorporated external features like time or location~\cite{adomavicius2010context}, but explicit user-declared context is more expressive and interpretable. When a system recommends a camera because a user identified as a professional photographer or filters out meat-based dishes for a vegetarian, the rationale becomes transparent and trustworthy.

Recent developments in LLM memory design, such as persistent memory in ChatGPT~\cite{openai2024memory} and episodic retention in Claude~\cite{anthropic2024memory}, illustrate alternative paradigms for maintaining evolving user context. For recommender systems, integrating structured user attributes (e.g., diet, expertise) with narrative memory extracted from feedback~\cite{he2024longtermmemory} supports both stable preference continuity and adaptive updates. As user context evolves, continual learning and adaptive replay~\cite{mi2020ader} ensure memory remains current and balanced between stability and flexibility. Overall, explicit user attributes form a bridge between personalization and alignment, grounding recommendation reasoning in user-declared context.

\subsection{Item-Level Comments and Quality Feedback}

Item-level feedback captures comprehensive and general assessments that combine a series of positive and negative reviews into neutral and objective quality features. A review such as “The performance is great for the price, but it overheats quickly and the fan is loud” simultaneously conveys item features regarding price and functions. Such holistic evaluations provide structured signals about trade-offs that go beyond isolated user-biased feedback. Review-aware methods have leveraged this information by constructing opinion graphs~\cite{cantador2021rating} or integrating aspect polarities into collaborative filtering~\cite{cui2024askat} to enhance explainability and precision.
Recent advances suggest that aggregating item-level narratives into structured knowledge representations can further strengthen alignment. Disentangling latent factors (e.g., price, usability, or quality) from reviews improves item embeddings and interpretability~\cite{ren2022disentangled}. Graph-based modeling~\cite{liu2021learning} and LLM-based feature induction can dynamically derive taxonomies of item attributes and summarize collective sentiment into machine-readable representations. Such integration grounds recommendations in the community’s textual feedback, enabling systems to justify suggestions based on aggregated user experience rather than opaque similarity scores.

In summary, item-level comments represent the richest and most integrative form of explicit feedback. They unify individual sentiments, contextual reasoning, and collective knowledge, offering both objective features and explainability. Harnessing these signals allows recommender systems to align better with user expectations while remaining transparent and trustworthy.

\section{LLM-Based RecSys Framework with Explicit User Feedback}

To transcend the limitations of traditional RecSys, we envision a multi-stage LLM-based framework that fully integrates explicit user feedback into the recommendation pipeline.
In this design, LLMs function as comprehension engines throughout the process, from interpreting user context to contextual reranking and explanation generation. The following subsections outline the key stages of this framework and their corresponding research challenges, forming a roadmap toward recommender systems that understand not only user behavior but also user rationale behind each action.

\subsection{Constructing and Updating User Profiles from Context}

The foundation of context-aware recommendation is a comprehensive user profile derived from explicit textual feedback rather than inferred behavior. This enables systems to model user rationale directly, creating faithful and interpretable representations of preference~\cite{bang2025llm}. LLMs can serve as agentic memory managers that continuously update user representations in natural language form, directly encoding rationales extracted from reviews or comments~\cite{shi2025personax,zhang2025personaagent}. For instance, a statement like “too sweet for my taste” can be directly translated into a structured dislike for “excessive sweetness,” eliminating ambiguity inherent in behavioral inference.

Structured representations further enhance the scalability and generalization of RecSys across domains. LLMs can distill user feedback into dynamic heterogeneous knowledge graphs, where typed edges capture evolving user–aspect relations over time~\cite{yang2023knowledge, zhang2025llminit}. For instance, a user’s positive remark about a concept or item may introduce a new node connected by a favor relation edge, while shifting preferences can lead to the relation type update or removal of existing interaction edges. Such text-to-graph mappings enable fine-grained tracking of preference dynamics, opening new research directions in preference evolution modeling and memory-based reasoning~\cite{bei2025graphs}.

\subsection{Context-Aware Retrieval and Interest Discovery}

The retrieval (candidate generation) stage is critical for both scalability and coverage in recommendation pipelines~\cite{hou2024bridging}. It filters billions of items into a small candidate set for fine ranking while ensuring high recall of relevant content. Since items missed at this stage cannot be recovered later, it defines the upper bound of the entire RecSys performance.
Retrieval in conventional RecSys focuses on semantic similarity, often missing the deeper rationale behind user preferences. Feedback-rewarded retrieval instead leverages the explicit user profiles, comprising their likes, dislikes, and real-time signals for candidate selection~\cite{adomavicius2010context,zhang2020explainable}. This shifts retrieval from pattern matching to rationale reasoning, enabling precise alignment between user intent and retrieved items.

An LLM can encode reasoning-infused embeddings that represent the user’s underlying motivation (e.g., preferring sci-fi for “philosophical themes” rather than “space battles” after reasoning) and apply them to dense retrieval~\cite{lewis2020retrieval,yang2025cold}. Additionally, at this stage, LLM-empowered RecSys can further conduct interest discovery to emerging or adjacent domains through agentic reasoning, where an LLM hypothesizes new interests by traversing a user’s knowledge graph or exploring multi-hop relations. Reinforcement learning techniques can reward discoveries that are novel yet consistent with explicit preferences, promoting serendipity and long-term engagement.

\subsection{Reranking and Rule Model Adaptation}

Once relevant candidates are retrieved and ranked, reranking refines the recommendation list to balance accuracy, diversity, and explainability~\cite{ziegler2005improving}. 
Traditional global rule models, which optimize uniform online engagement metrics, overlook individual rationales. Rule model adaptation could integrate explicit feedback into scoring through personalized contextual rules. Hard constraints (e.g., “allergic to peanuts”) can deterministically filter items, while soft preferences (e.g., “reduce caffeine”) modify scores proportionally. Hybrid frameworks that combine engagement likelihood with explicit alignment signals have shown improved personalization~\cite{burke2002hybrid,adomavicius2010context}. LLMs extend this capability by acting as evaluators that synthesize user rationales and item features to generate final scores~\cite{gu2024survey}. The final ranking thus reflects both predictive accuracy and user-specific refinement, ensuring transparent and trustworthy outcomes.

\subsection{Asynchronous Taxonomy and Tag Alignment}

A practical challenge is reconciling the deep understanding of LLMs with the latency constraints of real-time recommendation. To address this, we propose a research solution as an asynchronous mechanism based on tagging and taxonomy systems. LLMs operate in the backend to interpret user feedback and map it to an evolving taxonomy of semantic tags (e.g., \texttt{\#LightWeightPhone}, \texttt{\#LongBatteryLife} if one expresses satisfaction regarding the new phone features for light weight and long battery life). Thus, each tag is associated with an affinity score reflecting user sentiment~\cite{mcauley2015image,chen2017attentive}. When new feedback is provided, the LLM updates these scores asynchronously. The live retrieve, ranking, or reranking stages can then rely on precomputed tag features, enabling low-latency personalization real-time while informed by high-level reasoning at backend. This asynchronous mechanism bridges deep contextual comprehension of traditional RecSys and practical scalability of LLM-based RecSys.

\section{Conclusion and Future Vision}
Integrating explicit context feedback into recommender systems represents a paradigm shift from implicit inference to interactive preference alignment. By listening to the user’s own words, reviews, comments, and real-time feedback, systems can evolve from predictive models to conversational, adaptive assistants. This approach mitigates filter bubbles by introducing diverse, user-driven signals and enhances personalization by focusing on stated intent rather than historical patterns. Recent progress in review-based and LLM-driven recommendation already demonstrates gains in accuracy, interpretability, and transparency.
The next generation of RecSys should therefore treat explicit feedback as a primary data modality. With LLMs acting as interpreters of human rationale, future RecSys can engage in natural dialogue, ask clarifying questions, and justify recommendations in relatable terms. This shift not only improves personalization but also strengthens user agency and trust, turning recommendation from a passive process into an active collaboration between human and AI systems.

\bibliographystyle{IEEEtran}
\bibliography{main}

\end{document}